\documentclass[proof]{WileyASNA-v1}

\articletype{Article Type}%

\received{** *** 202*}
\revised{** *** 202*}
\accepted{** *** 202*}

\begin{document}

\title{Constraining stellar tidal quality factors from planet-induced stellar spin-up}

\author[1,2]{Nikoleta Ili\'c}
\author[1,2]{Katja Poppenhaeger}

\author[1,2]{Anna Barbara Queiroz}
\author[1]{Cristina Chiappini}
\authormark{Ili\'c \textsc{et al}}

\address[1]{\orgdiv{Cosmic Magnetic Fields}, \orgname{Leibniz Institute for Astrophysics Potsdam (AIP)}, \orgaddress{\state{Brandenburg}, \country{Germany}}}

\address[2]{\orgdiv{Institut f\"ur Physik und Astronomie}, \orgname{Universit\"at Potsdam}, \orgaddress{\state{Brandenburg}, \country{Germany}}}

\corres{Leibniz Institute for Astrophysics Potsdam (AIP), An der Sternwarte 16, 14482 Potsdam, Germany \\ \email{nilic@aip.de}}

\abstract{The dynamical evolution of tight star-planet systems is influenced by tidal interactions between the star and the planet, as was shown recently. The rate at which spins and orbits in such a system evolve depends on the stellar and planetary tidal dissipation efficiency. Here, we present a method to constrain the modified tidal quality factor $Q'_*$ of a planet-hosting star whose rotational evolution has been altered by its planet through angular momentum transfer from the planetary orbital motion to the rotation of the stellar convective zone. The altered rotation is estimated from an observed discrepancy of magnetic activity of the planet-hosting star and a coeval companion star, i.e.\ this method is applicable to star-planet systems with wide stellar companions. We give an example of the planet-hosting wide binary system HD189733 and find that the planet host's modified tidal quality factor is constrained to be $Q'_* \leq 2.33 \times 10^7$.}


\maketitle

\section{Introduction}

The theory of tidal evolution of gravitationally bound systems dates back over one century \citep{Laplace1829,Darwin1879,Hough1897} and has developed to a great extent since then \citep{Goldreich1966,Hubbard1974,Touma1994,Eggleton1998,Ogilvie2007}. It was applied when trying to explain the formation of the Solar System, as well as the various planet-moon configurations that it harbors \citep{Fish1967,Goldreich1968,Greenberg1973,Lin1986}. The evolution of close stellar binary systems was only understood after considering tidal interactions between the companion stars \citep{Zahn1977,Hut1981,Hurley2002}.
The consideration of tidal interactions between stars and planets began with the discovery of the first exoplanets \citep{Mayor1995,Rasio1996,Trilling1998,Cuntz2000}, and has since contributed to the understanding of the formation and evolution history of star-planet systems \citep{Bodenheimer2001,Ogilvie2004,Pont2009,Bolmont2016,Dawson2018}. 

Generally, two types of tides can occur in tidally interacting bodies: dynamical and equilibrium tides. Dynamical tides typically appear as inertial and gravity waves excited by the periodic tidal potential. Considering stars, inertial waves are excited in the convective zone if the tidal forcing frequency $\omega_{tide}$\footnote{The tidal forcing frequency is defined as $\omega_{tide} = 2|(\omega_{orb} - \omega_{spin})|$, where $\omega_{orb}$ is the orbital frequency of the perturber and $\omega_{spin}$ is the spin frequency of the host star.} is smaller than twice the spin frequency $\omega_{spin}$ of the star: $|\omega_{tide}| < 2|\omega_{spin}|$; their restorative force is the Coriolis force, while dissipation takes place due to the turbulent friction of convective motion \cite{Ogilvie2007}. Inertial waves excited in partly convective stars can reflect from the boundaries of the convective zone and result in enhanced tidal dissipation \citep{Ogilvie2013}. 

Tidally excited gravity waves can occur in the radiative zone of a star if the stellar oscillation frequency is close to the tidal forcing frequency; their restorative force is the buoyancy force in stably stratified regions, while the typical dissipation mechanism is radiative dampening \citep{Zahn1975}. If the tidally excited oscillations are in resonance with stellar gravity modes, tidal dissipation within the star can be enhanced through resonance locking \citep{Witte1999,Burkart2014,Fuller2017,Ma2021}. Another dissipation mechanism for tidally induced gravity waves in Sun-like stars is wave braking close to the center of the star where the non-linear component of the wave is dissipated \citep{Barker2010}. In the weakly non-linear regime gravity waves can lead to the excitation of daughter and granddaughter waves whose interaction can dissipate tidal energy as well \citep{Weinberg2012,Essick2016}.

Equilibrium tides are large-scale flows in the form of quasi-hydrostatic tidal bulges. They are formed in stellar convective envelopes under the influence of the periodic tidal potential of a perturber \citep{Darwin1879,Chandrasekhar1933,O'Leary1972,Soter1977,Zahn1977}. They occur in systems where the interacting bodies are not synchronous with each other's spin and orbital motion. This happens due to their viscous interiors that facilitate the dissipation of energy through convective turbulence and give rise to a net tidal torque.

Dynamical and equilibrium tides can occur simultaneously in a star: in partly convective stars, equilibrium tides occur as large-scale flows in the convective envelope, while dynamical tides in the form of gravity waves can occur in the radiative core; if the star is a fast rotator, dynamical tides in form of inertial waves can occur in the convective envelope as well. Considering an early-type star, with an outer radiative envelope and a convective core, gravity waves occur in the outer envelope, while excitation of weak equilibrium tides is possible in the convective core. The loss of tidal energy through the different mechanisms causes a net tidal torque between the interacting bodies, which changes the system's configuration on long timescales.

The rate of tidal evolution of a system depends, amongst others, on parameters that account for the efficiency of tidal dissipation in each interacting body. If we consider a star-planet system, these parameters are the tidal quality factor $Q_*$ and $Q_p$ of the star and of the planet, respectively. By definition, the tidal quality factor is the inverse of the phase lag angle between the tidal forcing potential and the tidal bulge \citep{Goldreich1966}. Another representation of tidal dissipation efficiency is in combination with the potential Love number of the second order: $Q' \approx Q/k_2$ \citep{Ogilvie2007}, which is the modified tidal quality factor. The Love number $k_2$ is a dimensionless measure of the interior density profile of the body. Therefore, the modified tidal quality factor $Q'$ accounts for the fact that astrophysical bodies are typically not homogeneous but have an internal density distribution. Quantifying the values of $Q'_*$ and $Q'_p$ can constrain the rate of tidal evolution of star-planet systems and their end configuration.

The value of the modified tidal quality factor is typically constrained through tidal interaction models. For a given set of initial conditions and a priori $Q'_*$ values, a sample of star-planet systems is evolved to their current age, and the resulting distribution of the systems is compared to the observed one. The adopted modified tidal quality factor is the one that reproduces the observed distribution. Some of the parameters used to constrain $Q'_*$ are the distributions of orbital eccentricities of planets \citep{Jackson2008,Bonomo2017}, obliquity distribution in star-planet systems \citep{Hansen2012}, distribution of extrasolar planets in circular orbits \citep{Penev2012}, and the orbital separation distribution of planets \citep{CollierCameron2018}, and was found to be in the range of $10^5 < Q'_* < 10^9$, where the lower value corresponds to higher dissipation efficiency and vice versa. Additionally, it was also found that the modified tidal quality factor might not be constant for a given sample but is rather a function of the tidal forcing frequency \citep{Ogilvie2007,Jackson2008, Penev2018,Patel2023}.

The study by \cite{Penev2018} constrained the stellar modified tidal quality factor of individual planet-hosting stars. They employed the tidal evolutionary model \texttt{POET} \citep{Penev2014} to simulate the tidal spin-orbit evolution of each system. It considers the effect of stellar evolution on the stellar spin: it takes into account evolving stellar structure, angular momentum loss due to the magnetized stellar wind, and core-envelope coupling which contributes to the exchange of angular momentum between envelope and core. In their study, the authors assume that the tides induced by the planet have spun up the star to the observed values within the estimated age of the system, and constrain $Q'_*$ by assuming that the initial spin rate of the given host star coincides with the spin rate of star in young open clusters (for the given mass range). They found a highly frequency-dependent dissipation, $Q' \sim P_{tide}^{ -3.1}$, for tidal periods between $0.5 < P_{tide} < 2$ days.

Aside from star-planet systems, binary stellar systems were used for estimating the efficiency of the undergoing tidal evolution as well. Single-lined spectroscopic binaries in open clusters were used for studying tidal circularization and synchronization rates \citep{Meibom2006}. The rate of tidal evolution was determined by comparing the observed rotational angular velocity of each Sun-like primary star to the angular velocity required for the star to reach (pseudo)synchronization. \cite{Khaliullin2007} and \cite{Khaliullin2010} utilized the apsidal motion of the binary orbital axes as indicators of the internal axial rotation of early-type (radiative) stars and tested the efficiency of radiative damping \citep{Zahn1975,Zahn1977} on tidal synchronization and circularization rate. More recent work utilized \texttt{POET}, adopted for systems on eccentric orbits, to constrain the modified tidal quality factor of a sample of Sun-like primary stars in eclipsing binaries by using their rotation periods and found a common value of $\log Q' = 7.818 \pm 0.035$ \citep{Patel2022}. A subsequent study explored the tidal frequency dependence of \textit{Q'} in Sun-like stars and parametrized Q' as a saturating power law in tidal frequency and obtained constraints using the rotation period of 70 eclipsing binaries observed by Kepler \citep{Patel2023}. 

This study presents an analytical method to calculate the modified tidal quality factor of slow-rotating planet-hosting stars under the equilibrium tide theory. For this, we use combined star-planet- and star-star systems, i.e. planet-hosting wide binary systems. We consider the tidal interactions taking place between a star and its planet, while we use the wide stellar companion as a reference point for the planet host to estimate the tidal impact it undergoes.

In section 2, we describe the method to constrain $Q'_*$ from observed activity differences in wide planet-hosting binary systems. In section 3, we demonstrate the methodology by applying the steps on the wide binary system HD~189733 and calculating the modified tidal quality factor of the planet-hosting star. A comparison with literature values for planet-hosting stars is given in Section 4, where we also discuss the method's applicability and issues regarding activity comparison, stellar spin-down, intrinsic activity variability, tidal evolution timescales, and uncertainties due to the employed activity-rotation relation. In section 5 we summarize the findings made in this study.

\section{Methodology}

The underlying assumption of the following analytical method relies on the results published by \cite{Poppenhaeger2014} and \cite{Ilic2022}. In particular, \cite{Ilic2022} showed that tidal interactions between a close-in planet and its host star can lead to a stellar spin-up. In their study, a sample of seventeen wide binary systems, where one of the stars has an orbiting planet, was examined. It was found that in systems where significant strength of tidal interactions between a star and its planet is expected, a significant difference between the activity levels of the two coeval stars is observed, the planet-hosting star being more X-ray luminous than its stellar companion. Since stellar X-ray luminosity is a magnetic activity indicator and empirical evidence points to the magnetic activity being driven by stellar rotation \citep{Pallavicini1981,Noyes1984,Pizzolato2003,Wright2011}, this result translates to an \textit{over-rotation} of planet-hosting stars. Therefore, we assume that the observed magnitude of over-activity and its interpretation as over-rotation contains information about the efficiency of tidal interactions between the planet and the host star. We outline a method here for how a measurement of over-activity can be used to constrain the modified tidal quality factor of the host star.

As a short overview, we use the following steps, which are explained in more detail in the following subsections: When a planet-hosting star in a wide binary system has observationally been determined to be over-active, we assign the difference in the stellar X-ray luminosities - corrected by the spectral type (SpT) difference - to the tidal impact an orbiting planet may have on the host star (see \citealt{Ilic2022}). By applying the activity-rotation relation for unsaturated main-sequence stars with an outer convective layer \citep{Wright2011}, we estimate the corresponding change in the rotation rate of the host star. Further, using the change of the host's rotation rate, we estimate the angular momentum $\Delta L$ that was exchanged between the planetary orbit and the stellar convective envelope, the change of the orbital semi-major axis $\Delta a$, and, lastly, the stellar modified tidal quality factor $Q'_*$.

\subsection{X-ray luminosity change}
\label{sec:activity_difference}

In the most general case, binary stars have different spectral types. Therefore, firstly the activity difference due to SpT difference has to be accounted for. \cite{Ilic2022} introduced a method where they employ the volume-limited sample of F/G-, K-, and M-type field stars in the solar neighborhood \citep{Schmitt2004}, observed with ROSAT, from which they assemble X-ray luminosity distribution function for each SpT (see Figure 3 in their study). The mean value of each distribution is shifted with respect to the other two, which is interpreted as the intrinsic difference in the activity level due to SpT difference. 

To account for the activity difference due to the SpT difference of binary stars, \cite{Ilic2022} compared the percentile positions of the stars, for which they employed the mean value and standard deviation of the respective luminosity distribution. If both stars have a similar state of their rotation and activity evolution, they should be found at similar percentiles of their respective distribution. However, if there is a significant difference in the percentile values, it is assumed to be due to the tidally induced spin-up and the subsequent X-ray luminosity increase of the planet host.

Similarly, here we assume that the difference in percentile values between the binary stars is rooted in the tidal impact of the planet and does not depend on the spectral type of the stars. Therefore, we assume that the planet-hosting star that does not experience any tidal impact should be found at the percentile position of the stellar companion, which we use to calculate the corresponding X-ray luminosity value. If we assume that the A component is the planet host and has SpT Y, while the B component has SpT Z, the expected {\it no-tidal-impact} X-ray luminosity of the A component is:

\begin{equation}
\label{eq:X-ray luminosity}
    L_{x(noTI)}(A) = f^{-1}_Y(f_Z(L_x(B))).
\end{equation}
Here, $f_Z$ is the X-ray luminosity distribution function for SpT Z and gives the percentile value for the given X-ray luminosity, while the $f^{-1}_Y$-function estimates the X-ray luminosity for the given percentile value in the X-ray luminosity distribution for SpT Y. For more details see Fig. \ref{fig:Lx_comp}.

One has to bear in mind that the field star sample obtained by \cite{Schmitt2004} consists of differently aged field stars found in various stages of their stellar evolution. Although main-sequence stars are particularly stable, their rotation rates and activity levels decrease with age, which introduces a broadening of the observed X-ray luminosity distribution. This implies that coeval stars with different spectral types are expected to have an activity difference that is less pronounced than suggested by the field stars sample. A preliminary analysis of wide binary systems indeed shows that the activity difference due to SpT difference is lower (Dsouza et al. in prep.). Therefore, basing an estimate of the tidally-driven activity difference between a planet host and its coeval companion on stellar samples with mixed ages such as the solar neighborhood leads most likely to an underestimate. Further work on wide binary systems can help make the estimates we outline here more precise.

\subsection{Rotation period change}
\label{sec:activity-rotation}

Having estimated the expected X-ray luminosity of the planet-hosting star in the absence of the tidal influence of its planet, we can further estimate the (slower) rotation period the host would have in this case. For this task, we employ the activity-rotation relation:

\begin{equation}
\label{eq:activity-rotation}
    \log R_x = \log C + \beta \log R_o.
\end{equation}

Here, the activity indicator is the ratio of the stellar X-ray luminosity to the bolometric luminosity $R_\mathrm{X} = L_\mathrm{X}/L_{\mathrm{bol}}$, the Rossby number is the ratio of the stellar rotation period to the convective turnover time $R_\mathrm{o} = P_{\mathrm{rot}}/\tau$, {\it C} is the proportionality constant, and $\beta$ is the power-law slope of the unsaturated regime. The slope value is adopted from the analysis by \cite{Wright2011} who determined $\beta = -2.7 \pm 0.13$. They also determined that the convective turnover time is a function of the stellar mass $M_*$:

\begin{equation}
    \log \tau = 1.16 - 1.49 \log (M_*/M_{\odot}) - 0.54 \log^2(M_*/M_{\odot}).
\end{equation}

This relationship is valid over the mass range between 0.09~-~1.36~$M_{\odot}$ and it engulfs main-sequence stars with a convective envelope.
 
Substituting $L_{\mathrm{x(noTI)}}(A)$ in Equation \ref{eq:activity-rotation} and assuming that the bolometric luminosity $L_{\mathrm{bol}}$ and convective turnover time $\tau$ of the planet host are unaffected by tidal interactions, the Rossby number $R_{\mathrm{o(noTI)}}$ and the rotation period $P_{\mathrm{rot(noTI)}}$ of the planet host in the case of no tidal impact (noTI) of the planet can be estimated. Here, it has to be considered that the stellar sample used to define the activity-rotation relation we employ introduces a significant scatter around the $\log R\mathrm{x}$ and $\log R\mathrm{o}$ relationship. We consider the scatter to be a source of uncertainty in the derivation of $R_{\mathrm{o(noTI)}}$, in addition to the measurement uncertainty of employed parameters. To quantify this uncertainty, we use the observed sample from \cite{Wright2011} to calculate by how much stars deviate from the activity-rotation relation in the unsaturated regime: $d\log Ro = \log \frac{P_{rot}}{\tau} - \frac{(\log R_x - \log C)}{\beta}$. We computed a standard deviation of $\sigma = 0.166$ dex for the entire sample, which we add to the uncertainty budget of $\log R_{\mathrm{o(noTI)}}$. Since the uncertainty of $P_{\mathrm{rot(noTI)}}$ is asymmetric in linear space (see e.g. Tab \ref{tab:TI_vs_noTI}), we use the mean uncertainty and assume error propagation to calculate the uncertainties of subsequent {\it no-tidal-impact} parameters.

\subsection{Angular momentum change}
\label{sec:ang_mom}

How the exchange of angular momentum between a planet's orbit and the stellar spin can influence the evolution of a star-planet system has been presented by \cite{Penev2012,Penev2014}: a reduction in the semi-major axis of the planet leads to the increased angular momentum of the host star. We make the simplifying assumption that a shrinking planetary orbit due to tidal star-planet interactions constitutes an angular momentum transfer onto the star, distributed over a certain amount of time, leaving other processes unchanged. In reality, there will be subsequent changes to the stellar wind and magnetic braking due to this process (see section~\ref{discussion_wind}). Therefore, the $\Delta L$ given below represents a lower limit of the actual angular momentum being exchanged.

Here we continue by estimating the expected change in angular momentum of the star from the estimated change in surface rotation derived in the previous step.
Stars are known to have differential rotation, both at the surface and in radial depth. However, in a modeling context, one often resorts to a simplification, where the convective envelope and the radiative core of low-mass main sequence stars are approximated to rotate as solid bodies at different rates \citep{Allain1998,Irwin2007,Bouvier2008,Denissenkov2010}. To quantify the amount of angular momentum transferred from the planet's orbital motion to the stellar spin, we only consider the rotation of the convective envelope and assume that the coupling timescale and angular momentum transport between the envelope and the core are not influenced by tidal interactions. 

The angular momentum of a spherical shell with a density profile $\rho(r)$ and the angular velocity $\omega_{\mathrm{surf}}$ is:

\begin{equation}
    L = 4 \pi \omega_{surf} \int_{R_{cz}}^{R_*} r^4 \rho(r) dr,
\end{equation}
where $R_*$ is the radius of the star, while $R_{\mathrm{cz}}$ is the height of the convective zone base in units of the stellar radius. Here, we assume the density profile of the Sun \citep{Bahcall2004}\footnote{The density profile was adopted from the following source: \url{http://www.sns.ias.edu/~jnb/SNdata/Export/BP2004/bp2004stdmodel.dat}.}, which, in the convective envelope, can be approximated by a second-order polynomial: $\rho (r) = a\,r^2+b\,r+c$, where $a = 4.753\times 10^{-15}\,\mathrm{kg/m^5}$, $b = -6.571\times 10^{-6}\,\mathrm{kg/m^4}$, and $c = 2276\,\mathrm{kg/m^3}$.

The lower limit (LL) of the exchanged angular momentum of the stellar convective envelope due to tidal interactions with an orbiting planet is:

\begin{equation}
\label{eq:dL}
    \Delta L_{LL} = 8 \pi^2 \left( \frac{1}{P_{rot}}-\frac{1}{P_{rot(noTI)}} \right) \int_{R_{cz}}^{R_*} r^4 \rho(r) dr
\end{equation}

\subsection{Semi-major axis change}

Having estimated $\Delta L_{LL}$, we can constrain the change of the semi-major axis due to tidal interactions:

\begin{equation}
\label{eq:da}
\Delta a_{LL} =  \left(\frac{\Delta L_{LL}}{M_pM_*\sqrt{\frac{G}{M_p+M_*}}}+\sqrt{a}\right)^2-a.  
\end{equation}

Here, $M_\mathrm{p}$ and $M_*$ are the masses of the planet and the star, {\it a} is the observed semi-major axis of the planet's orbit, and {\it G} is the gravitational constant.

The above equation was derived by integrating Equation 2 given by \cite{Penev2012,Penev2014} over the boundaries from $L \rightarrow L+\Delta L_{LL}$ for the stellar angular momentum and from $a+\Delta~a_{LL}~\rightarrow~a$ for the orbital semi-major axis: stellar rotation gains angular momentum, while the semi-major axis of the planetary orbit is reduced due to the loss of angular momentum. 

It is assumed that the angular momentum contribution from the planet's rotation is small and that the tides raised on the planet force it into synchronization with the orbital motion quite early in the system's evolution. Therefore, it is assumed that the angular momentum is only exchanged between the planetary orbit and the stellar spin. Also, by using the above equation, we assume that the planetary orbit is circular and aligned with the stellar equator. Therefore, the angular momentum loss only reduces the orbital semi-major axis, not the orbital eccentricity or inclination.

\subsection{The modified tidal quality factor $Q'_*$}

Finally, to estimate the upper limit (UL) of the modified tidal quality factor $Q'_*$, we employ the equation for tidal evolution of the orbital semi-major axis given as Equation 1 by \cite{Penev2012,Penev2014}. Here, we integrate over the time the planet migrates inward - $T_{\mathrm{in}} \rightarrow T_{\mathrm{sys}}$ - and the semi-major axis change - $a+\Delta a_{LL} \rightarrow a$:

\begin{equation}
\label{eq:Q}
    Q'_{*,UL} = -\frac{117}{4}\Delta T R_*^5 M_p \sqrt{\frac{G}{M_*}} \left(a^{\frac{13}{2}}-(a+\Delta a_{LL})^{\frac{13}{2}}\right)^{-1}.
\end{equation}

Here, $\Delta T = T_{\mathrm{in}}-T_{\mathrm{sys}}$, where $T_{\mathrm{in}}$ and $T_{\mathrm{sys}}$ are the time after formation when the planet started migrating inward and the age of the system, respectively. The factor \textit{-1} stems from $\mathrm{sign}(\omega_{\mathrm{surf}}-\omega_{\mathrm{orb}})$ of the original formula, where $\mathrm{\omega_{surf}}$ and $\mathrm{\omega_{orb}}$ are the angular velocities of the stellar surface and the planet in its orbit, respectively. It indicates the direction of angular momentum exchange: if the orbital angular velocity is higher than the angular velocity of the stellar surface, the angular momentum is transferred from the planet's orbit to the star. This is the case when the planet is found inside the system's co-rotation radius, the distance from the star where the planet's orbital rate and the stellar rotation rate are equal.

Depending on the initial conditions of the system, e.g. the initial stellar spin and the orbital distance of the planet after the protoplanetary disk has cleared, the outward or inward migration of the planet due to tidal interactions occurs. Typically, given that a newly formed star experiences a period of contraction and spin-up, the co-rotation radius is close to the star, while the planet is most likely located at a greater distance. Therefore, angular momentum is transferred from the stellar spin to the orbit, and the planet migrates outward. 

After the stellar wind takes over the dominance over the stellar rotation evolution, the co-rotation radius gradually increases, crossing the planet's orbit and leading to the inward migration of the planet. The timescale of the described process is typically a few $\times 10^7$ yr \citep{Bolmont2012}, however, depends on the initial stellar spin and planet location, the efficiency of the stellar wind and tidal dissipation, as well as on the stellar and planetary mass. Hereafter, tidal interactions and magnetic braking dictate the evolution and stability of the star-planet system.

As in the previous step, a circular orbit and synchronization between the planet's rotation and orbital motion are assumed.

\section{Example: the planet-hosting wide binary system HD~189733}

Here we present the application of the described method to the Hot Jupiter-hosting system HD~189733~AbB.
\subsection{The system}

The system HD~189733~AbB is composed of an early-K dwarf and a mid-M dwarf, that are separated by at least $\approx 220$ AU, which makes it a wide stellar binary system \citep{Bakos2006,Mugrauer2019}. The primary star is orbited by a Hot Jupiter-type planet with an orbital period of $\approx 2.2$ days \citep{Bouchy2005}. The stellar parameters of the host and the relevant planetary parameters are given in Table \ref{tab:HD189733}. The planet has a low-eccentricity orbit that is aligned with the stellar equator.

\begin{table}
	\centering
	\caption{Stellar and planetary parameters of the star-planet system HD~189733~Ab used and derived in this study. We have used the following references: \cite{Addison2019} as A19, \cite{GaiaDR3} as GaiaDR3, \cite{Ilic2022} as I22, and \cite{Henry2008} as H\&W08.}
	\label{tab:HD189733}
	\begin{tabular}{lcccc} 
        \hline
        Parameter & Value & Reference \\
        \hline
		$\mathrm{M_*}$ [M$_{\odot}$] & $0.81 \pm 0.03$ & A19 \\[0.1cm]
        $\mathrm{R_*}$ [R$_{\odot}$] & $0.78 \pm 0.01$ & A19 \\[0.1cm]
        $\mathrm{T_{eff}}$ [K] & $5053_{-45}^{+46}$ & A19 \\[0.1cm]
        $\mathrm{\log g}$ & $4.563_{-0.020}^{+0.021}$ & A19 \\[0.1cm]
        [Fe/H] & $-0.003_{-0.029}^{+0.031}$ & A19 \\[0.1cm]
        $\mathrm{A_v}$ & $0.127^{+0.059}_{-0.058}$ & A19 \\[0.1cm]
        $\pi$ [mas] & $50.57 \pm 0.02$ & GaiaDR3\\[0.1cm]
        G [mag] & $7.4284 \pm 0.0003$ & GaiaDR3\\[0.1cm]
        Age [Gyr] & $6.97 \pm 1.02$ & this work\\[0.1cm]
        $\mathrm{L_{bol}}$ [erg/s]  & $(1.358 \pm 0.053) \times 10^{33}$ & A19\\[0.1cm]
        $\mathrm{L_x}$ [erg/s] & $(1.296 \pm 0.016) \times 10^{28}$ & I22 \\[0.1cm]
        $\mathrm{R_x}$ & $(9.5 \pm 0.4)\times 10^{-6}$ & this work \\[0.1cm]
        $\mathrm{P_{rot}}$ [day] & $11.953 \pm 0.009$ & H\&W08 \\[0.1cm]
        $\tau$ [day] & $19.8 \pm 1.2$ & this work\\[0.1cm]
        $\mathrm{R_o}$ & $0.605 \pm 0.044$ & this work\\[0.1cm]
        $\mathrm{M_p}$ [$\mathrm{M_{Jup}}$] & $1.130^{+0.047}_{-0.045}$ & A19\\[0.1cm]
        $\mathrm{R_p}$ [$\mathrm{R_{Jup}}$] & $1.142^{+0.036}_{-0.034}$ & A19\\[0.1cm]
        $\mathrm{P_{orb}}$ [day] & $2.2185788^{+0.0000091}_{-0.0000076}$ & A19\\[0.1cm]
        a [AU] & $0.03098^{+0.00043}_{-0.00039}$ & A19\\[0.1cm]
        e & $0.024^{+0.026}_{-0.014}$ & A19\\[0.1cm]
        i $\mathrm{[^o]}$ & $85.27^{+0.24}_{-0.23}$ & A19\\[0.1cm]
		\hline
	\end{tabular}
\end{table}

The system was observed in the X-ray regime by the Chandra X-ray Observatory and the XMM-Newton Space Telescope. Since the components have an angular separation of $\approx 11^{ \prime \prime}$, only \textit{Chandra} can resolve the two components. Here, we adopt the X-ray luminosity values estimated by \cite{Ilic2022}, who analyzed the {\it Chandra} observations and provided the percentile values of the stars in their respective X-ray luminosity distributions (see Section \ref{sec:activity_difference} for more details). They found that HD~189733~A has an X-ray luminosity of $ L_\mathrm{x}(A)~=~1.3~\times~10^{28}$~erg/s, corresponding to the $\mathrm{78^{th}}$ percentile in the K-dwarf distribution, while HD~189733~B has an X-ray luminosity of $L_\mathrm{x}(B)~=~7.4~\times~10^{26}$~erg/s, corresponding to the $\mathrm{48^{th}}$ percentile in the M-dwarf distribution. 

\begin{figure}
	\includegraphics[width=\columnwidth]{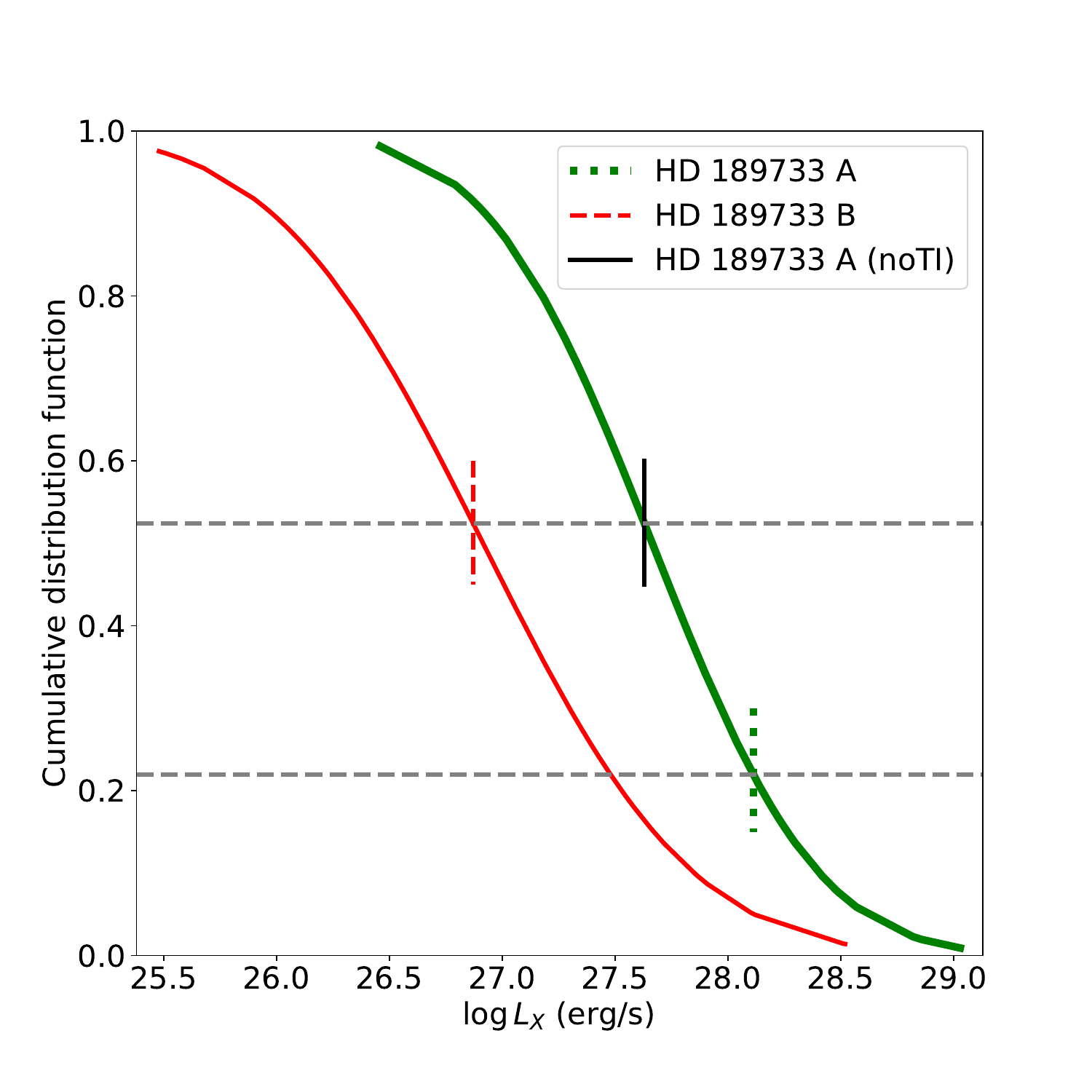}
    \caption{The cumulative X-ray luminosity distribution function of K- and M-dwarfs in the solar neighborhood as thin red and thick green solid curves, respectively. The green dotted and red dashed lines show the position of HD~189733~A and B, respectively, while the black solid line indicates the position of HD~189733~A in the {\it no-tidal-impact} regime.}
    \label{fig:Lx_comp}
\end{figure}

Figure \ref{fig:Lx_comp} shows the cumulative distribution functions of the X-ray luminosities for K- and M-dwarf stars in the solar neighborhood in the energy range 0.2-2.0 keV. The green dotted and red dashed lines show the position of HD~189733~A and B, respectively, while the black solid line indicates the position of HD~189733~A if the orbiting Hot Jupiter did not have any tidal impact on the star (see Section \ref{sec:app_method} for more details).

\subsection{Age of the planet host}

To estimate the value of the modified tidal quality factor $Q'_*$ for the planet host, we need knowledge of the system's age. The time interval in which the planet has migrated inward and exchanged angular momentum with the star will constrain the efficiency of tidal dissipation within the system: the less time the system needs to evolve to the current state, the more efficient the tidal dissipation seems to be (see Eq. \ref{eq:Q}).

The age of HD~189733~A was estimated using stellar evolutionary models \citep{Torres2008,Ghezzi2010,Southworth2010,Maxted2015} and discussed in the context of the primary's and secondary's X-ray luminosity \citep{Sanz-Forcada2010,Poppenhaeger2013, Poppenhaeger2014}. Since tidal interactions with a Hot Jupiter-type planet have shown to be able to alter the stellar rotation rate and X-ray emission, i.e.\ making the host appear younger, we will not consider the ages estimated by using the primary's X-ray luminosity. The ages estimated with stellar evolutionary models range from 1-7 Gyr for HD~189733~A, with an uncertainty budget close to 100\%.

To obtain a better constraint on the age of the planetary host, we use the isochrone-matching code \texttt{StarHorse}, a Bayesian tool for estimating stellar parameters, including age, based on photometric, spectroscopic and astrometric data \cite{Queiroz2018,Santiago2016}. As input parameters for HD~189733~A, we apply the {\it Gaia} DR3 {\it G} magnitude and parallax, mass, radius, bolometric luminosity, effective temperature, surface gravity, metallicity, and V-band extinction given in Table \ref{tab:HD189733}. With this tool, we estimated the age of HD~189733~A to be $6.97 \pm 1.02$ Gyr, which indicates a system older than the Sun. The age of the host is now better constrained with an uncertainty of less than 15\%.

\subsection{$Q'_*$ of HD~189733~A}
\label{sec:app_method}

Having the observed X-ray luminosities of stellar components of HD~189733, we can estimate the X-ray luminosity of the host if the orbiting planet did not have any tidal impact using Eq. \ref{eq:X-ray luminosity}. This corresponds to an X-ray luminosity of $L_{\mathrm{X(noTI)}}(A)~=~(4.26\pm0.21)~\times~10^{27}$~erg/s. Here, we assumed the luminosity uncertainty of 5\%, corresponding to the X-ray luminosity uncertainty of the B component. 

\begin{figure}
	\includegraphics[width=\columnwidth]{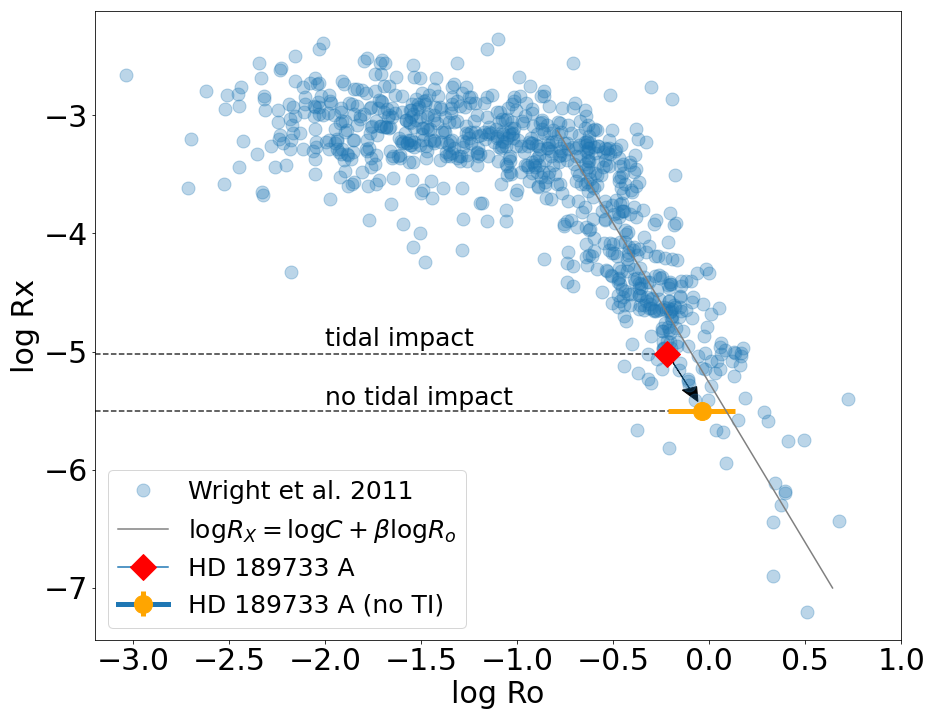}
    \caption{The stellar sample used by \protect\cite{Wright2011} and the activity-rotation relation fitted for the bulk of the unsaturated regime with the slope $\mathrm{\beta = -2.7}$ (grey line). The planet host, as observed, is shown with the red diamond symbol, while the filled orange circle is the hypothetical {\it no-tidal-impact} position of HD~189733~A. The positional change of HD~189733~A due to the absence of tidal interactions is indicated by the black arrow.}
    \label{fig:wright+2011_hd189733}
\end{figure}

By applying Equation \ref{eq:activity-rotation}, and assuming that the bolometric luminosity $L_{\mathrm{bol}}$ and the convective turnover time $\tau$ are not affected by tidal interactions, the planet host would have a rotation period $P_{\mathrm{rot(noTI)}} = 18.04_{-5.79}^{+8.53}$~days. Table \ref{tab:TI_vs_noTI} summarizes the activity and rotation parameters of the planet-hosting star in the regime of tidal impact (TI) and no tidal impact (noTI) from the planet. In addition, Figure \ref{fig:wright+2011_hd189733} shows the stellar sample used by \cite{Wright2011} and HD~189733~A. The difference in the position of the planet host in the two regimes shows the significance of tidal interactions in the evolution of star-planet systems.

\begin{table}
    \centering
    \caption{Parameters of HD~189733~A observed and estimated in the cases when the orbiting planet has a tidal impact (TI) and when it does not (no TI). The uncertainty of $\log R_{\mathrm{o}}$ for the noTI regime includes the contribution due to the stellar scatter around the activity-rotation relation.}
    \label{tab:TI_vs_noTI}
	\begin{tabular}{l|cc} 
	
        \hline
	  Parameter & TI & noTI \\
        \hline
        $\mathrm{L_x}$ [erg/s] & $(1.296 \pm 0.016) \times 10^{28}$ & $(4.263 \pm 0.213) \times 10^{27}$\\[0.2cm]
        $\mathrm{\log R_x}$ & $-5.020 \pm 0.018$ & $-5.503 \pm 0.028$\\[0.2cm]
        $\mathrm{\log R_o}$ & $-0.218 \pm 0.025$ & $-0.039 \pm 0.174$\\[0.2cm]
        $\mathrm{P_{rot}}$ [day] & $11.953 \pm 0.009$ & $18.04_{-5.79}^{+8.53}$\\
        \hline
    \end{tabular}
\end{table}

The lower limit of angular momentum transferred from the planet's orbit to the stellar convective envelope given by Equation \ref{eq:dL}, the minimal change of the orbital semi-major axis given by Equation \ref{eq:da}, and the maximal stellar modified tidal quality factor given by Equation \ref{eq:Q} are summarized in Table \ref{tab:change_parameters}. For Eq. \ref{eq:dL}, we estimated the height of the convective envelope base to be $R_{\mathrm{cz}} = 0.677\,R_*$ using the K-dwarf stellar sample by \cite{Wright2011}. The modified tidal quality factor of HD~189733~A is $Q'_* \leq 2.33\times 10^7$. The uncertainties given in Table \ref{tab:change_parameters} are large, mainly driven by the scatter in the rotation-activity relationship. The derived order of magnitude of $Q'_*$ indicates a moderately, potentially highly efficient tidal dissipation in the stellar interior.

\begin{table}
    \centering
    \caption{Parameters derived with our analytical method and used to deduce the tidal dissipation efficiency of HD~189733~A.}
    \label{tab:change_parameters}
	\begin{tabular}{lll} 
        \hline
	Parameter & Value & Uncertainty\\
        \hline
	$\mathrm{\Delta L_{LL}}$ [$\mathrm{kg\,m^2/s}$] & $4.9 \times 10^{40}$ & $3.8 \times 10^{40}$ \\
        $\mathrm{\Delta a_{LL}}$ [$\mathrm{AU}$] & 0.0020 & 0.0016\\
        $\mathrm{Q'_{*,UL}}$ & $1.19 \times 10^7$ & $1.14 \times 10^7$\\

        \hline
    \end{tabular}
\end{table}

\section{Discussion}

\subsection{The derived $Q'_*$ constraint in context}

We have introduced a method to analytically, using X-ray observations of a planet-hosting wide binary system, constrain the modified tidal quality factor $Q'_*$ of the host star which tidally interacts with its {\it close-in massive} planet. The main assumption is that if the planet host appears to be more active than its stellar companion, the excess activity is due to tidal spin-up or at least a slow-down of the process of magnetic braking due to the magnetized stellar wind. Our constraints, when used for a single star-planet system, are quite loose; however, with improvements to stellar rotation-activity relationships as well as with applications to larger samples of star-planet systems for stars of similar spectral types, one can expect to derive tighter constraints.

Until now, the usual approach in estimating the stellar $Q'$ was to employ tidal evolution models and try to reproduce the distribution of observed star-planet parameters. The study of \cite{Jackson2008} determined $Q'_* \approx 10^{5.5}$ by reproducing low-eccentricity values for exoplanets closer than $a=0.2$~AU to their host. On the other hand, \cite{Hansen2012} estimated the stellar modified tidal quality factor of $Q'_* \approx 10^7 - 10^8$ by reproducing the orbital orientation of systems with $P_{\mathrm{orb}} < 3$~days. They also found indications that $Q'_*$ depends on the thickness of the stellar convective envelope. Furthermore, \cite{Bonomo2017} estimated $Q'_* \geq 10^6 - 10^7$ by comparing circularization timescales with stellar ages for planets within $a < 0.05$~AU, and \cite{Penev2018} found that the observed rotation rate of planet hosts can be explained if tidal dissipation depends on the forcing frequency with $Q'_* \approx 10^5$ at $0.5\,\mathrm{day}^{-1}$ to $Q'_* \approx 10^7$ at $2\,\mathrm{day}^{-1}$ for $P_{\mathrm{orb}} < 3.5$~days.

The exoplanet HD~189733~Ab has an orbital period of $P_{\mathrm{orb}}=2.2$~days and is separated from the host star by $a = 0.03$~AU. Having an eccentricity of $e = 0.024$ and an inclination of $i~=~85.3^{\circ}$, its orbit is almost circular and aligned to the host's equator. With $Q'_* \leq 2.33 \times 10^7$, the tidal dissipation efficiency of HD~189733~A as a Hot Jupiter-hosting star, determined using our analytical method is within the estimates made by reproducing distributions of star-planet systems with tidal models. The tidal forcing frequency in the HD~189733~A star-planet system is $\omega_{tide} \approx 4.5\,\mathrm{day^{-1}}$, which shows that the constrain we put on the $Q'_*$-value corresponds with the conclusion made by \cite{Penev2018}.

\subsection{Applicability of the analytical method}

The uncertainties of the parameters given in Table \ref{tab:change_parameters} are quite significant. The largest contributor to their uncertainties is the stellar scatter of the activity-rotation relation. Considering the study by \cite{Ilic2022}, the planet-hosting wide binary HD~189733 has the largest currently observed activity difference between two stellar components, i.e.\ already presenting a favorable case.

For the future, an analysis of a sample of planet-hosting wide binary systems with similar host and planetary orbit parameters should yield a well-constrained modified tidal quality factor for a specific star-planet configuration. Otherwise, analyzing planet hosts with different spectral types could yield a dependence of the $Q'_*$-value with stellar parameters like the convective envelope depth, as was suggested by \cite{Hansen2012}. Also, estimating the forcing frequency $\omega_{tide}$ for each such star-planet system could contribute to a better understanding of the relationship between $Q'_*$ and $\omega_{tide}$.

The binary sample that could be employed in this endeavor can be determined from the work done by \cite{Mugrauer2019}. The author uses Gaia DR2 to find wide stellar companions of 1300 planet-hosting stars within 500 pc of the Sun. Out of these stars, 176 had one stellar companion, while 27 were found in hierarchical triples. \cite{Ilic2022} used the sample and found 17 resolved wide binary systems, out of which 11 have a determined rotation period of the host star (a parameter needed for estimating $Q'_*$ with our method). Further, three systems have a Hot Jupiter orbiting the primary star and both stellar components are detected, while three other systems have a Hot Jupiter orbiting the primary star, but the stellar companion is not detected in X-rays. Therefore, currently, for four planet-hosting stars (including HD~189733~A) we can constrain $Q'_*$, and for three host stars, we can calculate the upper limit. \cite{Ilic2022} searched for serendipitous detections in archival observations of the Chandra- and XMM-Newton space telescopes. However, with dedicated observations, up to 30 resolved, Hot Jupiter-hosting wide binary systems, not previously observed, can be found based on the sample published by \cite{Mugrauer2019}. As filters for detectability of the systems, we considered the distance to the systems and instrumental spatial resolution power: up to 200 pc and above $15^{''}$ for observations with XMM-Newton space telescope, and up to 100 pc and angular separation between 3 and $15^{''}$ (to avoid overlap with systems observable with XMM-Newton) for the Chandra space telescope. In both cases, we considered a separation of at least 100 AU between the stars to avoid perturbation of the star-planet system by the stellar companion \citep{Desidera2007}. One additional constraint that can reduce the number of potential systems is the availability of rotation periods of planet-hosting stars.

\subsection{Caveats}

\subsubsection{Comparing activity levels of coeval stars}

In step 1 of the analytical method, described in Section \ref{sec:activity_difference}, we assume that similar coeval stars residing in a wide binary system will evolve as single stars, and would have similar activity levels and rotation rates if there is no close companion with whom they can tidally interact. This assumption is based on the behavior of stars found in the unsaturated regime of the activity-rotation relation. Figure \ref{fig:wright+2011_hd189733} shows that the ratio of the X-ray luminosity to the bolometric luminosity of stars ($R_x$) decreases with increasing Rossby number $R_o$, which is the ratio of the stellar rotation period and its convective turnover time. This relation points to the fact that as stars lose momentum and spin down due to the magnetized stellar wind, their magnetic activity will decrease as well. Therefore, the coupled evolution of stellar rotation and activity indicates that two similar stars born at the same time will on average experience similar spin-down rates, and be found in a similar evolutionary stage. Subsequently, they will appear to have similar X-ray luminosities.

This conclusion might seem counterintuitive if we take a look at the X-ray luminosity distribution in Fig. \ref{fig:Lx_comp}: there we see a wide spread of stars with the same spectral type, ranging over three orders of magnitude. However, shown is a volume-limited stellar sample found within 10 pc of the Sun. These field stars cover a range of ages and have experienced different angular momentum loss rates, which led to various rotation rates and activity levels, as is anticipated from the activity-rotation relation. Therefore, it is reasonable to assume that the activity levels of two coeval stars of the same SpT are more consistent with each other than what is observed between two filed stars that do not necessarily share the same age.

Strictly speaking, when equalizing the percentile value of the planet host with no tidal impact to the percentile value of its planet-free stellar companion, we are assuming that their activity levels are \textit{the same}. The assumption that the stars would have the same activity level if they are of the same spectral type, or be at the same percentile if they have different spectral types is a simplification of the issue at hand. One cause for stars having different activity levels can be short- and long-term variability, which we discuss in Sections \ref{sec:variability}; e.g. it was found that single M dwarf stars can vary by a factor of 2 in time-averaged X-ray data \citep{Magaudda2022,Ilic2023}. However, until now no studies have tackled the problem of activity variability between Sun-like coeval stars. Therefore, we are left with the assumption made in step 1 of the analytical method.

Finally, we point out that the largest contribution to the uncertainties of the tidal-interaction parameters originates from the scatter of the stellar sample employed in the activity-rotation relation (see Section \ref{sec:activity-rotation}). Therefore, we consider that the difference in the activity level between the stellar companion and the host star in the no-tidal-impact regime not covered by our normalization procedure (if there is any), is covered by those uncertainties.

\subsubsection{The impact of magnetic braking}\label{discussion_wind}

The way we treated the angular momentum exchange between planetary orbit and host star spin in section \ref{sec:ang_mom} amounts to an angular momentum dump stretched over a certain period while assuming that all other spin-down relevant processes stay the same. However, stars with higher activity levels are expected to have more efficient magnetic braking due to their wind properties (e.g. \citealt{Mestel1968,Mestel1987}). This indicates that the estimated amount of angular momentum exchanged between the star and the planet, derived from the X-ray luminosity difference between the stellar components, is a lower limit since the host star likely experiences a higher magnetic braking rate than its stellar companion. Following this conclusion, the estimated semi-major axis change represents a lower limit as well, while the estimated value of the modified tidal quality factor is an upper limit, as shown in Table \ref{tab:change_parameters}.

To minimize the effect of different spin-down rates between stellar companions, additional caution is needed when comparing stars with different internal structures, e.g. binary companions with an outer radiative and convective envelope, or partly and fully convective stars: in these kinds of pairs, the mechanisms contributing to their angular momentum loss might be different. We, therefore, consider our described method to apply to pairs of stars with a convective envelope and a radiative core.

\subsubsection{Stellar activity variability}
\label{sec:variability}

The stellar activity level can vary on different timescales. This will impact the observed activity difference between stellar components. The activity variability intrinsic to a star, not caused by external sources, can have various origins and has the form of stochastic variability on short timescales, or of activity cycles on longer timescales. Given the usual observation time of 30-100 ks in the X-ray domain, the stochastic variability of a star is to some degree accounted for when estimating the X-ray luminosity from one observation run. To also cover the activity variability due to an activity cycle, if the considered stars experience any at all, the binary system should be observed in different epochs, spanning over several years. This approach will cover the different activity regimes and will provide the average activity level of both stars. Comparing these activity levels will yield the tidal impact of an orbiting planet, with minimal contribution by the intrinsic activity variability of individual stars.

\subsubsection{Tidal evolution timescales}

The most general orbit is eccentric and inclined to the stellar equatorial plane. We have assumed that the timescales for orbital circularization and alignment are shorter than the stellar age. This might apply to the star-planet system HD~189733~A since the planet's orbit has a very small eccentricity and an almost aligned orbit. These characteristics however do not constrain the timescales or the amount of angular momentum that was lost to the stellar spin in the process of circularization or alignment. Therefore, the stellar angular momentum gain from the reduction of the orbital semi-major axis assumed in our methodology provides, strictly speaking, only a lower limit on the modified tidal quality factor $Q'_*$, meaning that the host star might be less efficient in tidal dissipation.

An additional issue related to the orbital timescales is the planet's spin-to-orbit synchronization time. Again, by neglecting the angular momentum that was exchanged between the planet's spin and the orbit, we are breaking the angular momentum conservation law. However, the angular momentum of the planet is orders of magnitude smaller than that of the star and contributes to the overall angular momentum budget of the star-planet system only marginally. If the planet's spin and orbit are asynchronous and the orbit is eccentric, the synchronization process can contribute to the circularization of the orbit if the planet's spin rate is lower than the orbital rate \citep{Dobbs-Dixon2004}. This alleviates the significance of an eccentric orbit and an asynchronous planet spin in the evolution of the stellar spin; however, the synchronization timescale for this scenario depends on the tidal dissipation efficiency of the planet \citep{Matsumura2010}, which constrain is out of the scope of this paper.

\section{Summary}

To better understand the tidal evolution of star-planet systems, the knowledge of their tidal dissipation efficiencies in the form of the tidal quality factor $Q_*$ for stars and $Q_\mathrm{p}$ for planets is needed. Here, we introduce an analytical method to estimate the stellar modified tidal quality factor $Q'_* \approx Q_*/k_\mathrm{2}$, which is the ratio of the tidal quality factor and the second-order Love number for planet-hosting stars that reside in wide stellar binary systems.

We demonstrate the analytical method on the planet-hosting wide binary system HD~189733. The planet-hosting star has a modified tidal quality factor of $Q'_* \leq 2.33\times 10^7$, consistent with results for short orbital period systems. The uncertainty budget, stemming mostly from the stellar sample of the employed activity-rotation relation, suggests that a better constraint of the $Q'_*$-factor can be achieved on a sample of planet-hosting wide binary systems with similar star-planet configurations.

In general, we recommend considering binary systems where stellar components have a convective envelope and a radiative core, and, optimally, have the same SpT to minimize the impact of the different rates of magnetic braking between the two components. Further, assuming that stars have activity cycles, the best practice in estimating the average X-ray luminosities of stellar components is to use X-ray observations that span over several years, optimally covering different phases of the activity cycles. Lastly, this method should be applied to star-planet systems where the planet's orbit has low eccentricity and is aligned with the stellar equatorial plane, as well as where the stellar rotation period is at least double the orbital rotation period: $P_{rot} \geq 2P_{orb}$, to have no excitation of dynamical tides in form of inertial waves.

\section*{Acknowledgments}

We thank the anonymous reviewer for the valuable comments. NI and KP acknowledge support from the German Leibniz-Gemeinschaft under project number P67/2018.
This research made use of Astropy,\footnote{\url{http://www.astropy.org}} a community-developed core Python package for Astronomy \citep{Astropy2013, Astropy2018}. 
This work has made use of data from the European Space Agency (ESA) mission
{\it Gaia}\footnote{\url{https://www.cosmos.esa.int/gaia}}, processed by the {\it Gaia}
Data Processing and Analysis Consortium (DPAC\footnote{\url{https://www.cosmos.esa.int/web/gaia/dpac/consortium}}). Funding for the DPAC has been provided by national institutions, in particular, the institutions
participating in the {\it Gaia} Multilateral Agreement.











\nocite{*}
\bibliography{Wiley-ASNA}%



\end{document}